\documentclass{icrc2009}

\usepackage{graphicx}   
\usepackage[caption=false]{caption}    
\usepackage[font=footnotesize]{subfig} 
\usepackage{fixltx2e}
\usepackage{url}
\usepackage{amssymb}

\newcommand{\shorttitle}[1]%
{\markboth{Proceedings of the 31\MakeLowercase{$^{st}$} ICRC, {\L}\'{o}d\'{z} 2009}{#1} }
\newcommand{\etal}{\MakeLowercase{\textit{et al. }}} 

\begin{document}
\title{Search for an extended emission around blazars \\
with the MAGIC telescope.}

\author{\IEEEauthorblockN{Julian Sitarek\IEEEauthorrefmark{1}\IEEEauthorrefmark{2},
			  Razmik Mirzoyan\IEEEauthorrefmark{1} 
                          for the MAGIC Collaboration}
                            \\
\IEEEauthorblockA{\IEEEauthorrefmark{1} {Max-Planck-Institut f\"ur Physik, D-80805 M\"unchen, Germany}}
\IEEEauthorblockA{\IEEEauthorrefmark{2} {University of \L\'od\'z, PL-90236 \L\'od\'z, Poland}}}


\shorttitle{J.Sitarek \etal Search for AGN halo.}
\maketitle

\begin{abstract}
Very high energy gamma rays coming from extra-galactic sources can interact with intergalactic radiation fields. This process may result in electromagnetic cascades with the following cycle: the production of electron-positron pairs and then secondary gamma-rays due to inverse Compton scattering. Since electrons and positrons will be scattered in the intergalactic magnetic field, under certain conditions their radiation may be redirected towards the observer. Thus one can anticipate that the secondary gamma-ray emission may produce an apparent extended halo around the source.

MAGIC is an Imaging Atmospheric Cerenkov Telescope located on Canary island of La Palma at Roque de los Muchachos Observatory (2200 m.a.s.l). 
Various source sizes and extended emission profiles within $1^\circ$ diameter have been studied by using dedicated Monte Carlo simulations for the MAGIC telescope. 
We present results of the study of a possible extended emission for Mrk 421 and Mrk501 done with the MAGIC telescope.
  \end{abstract}

 
\section{Introduction}

Blazars are well-known extragalactic emitters of Very High Energy (VHE) $\gamma$-rays.
Their radiation transverses large distances in extragalactic space filled with CMB (Cosmic Microwave Background) and EBL (Extragalactic Background Light) photons and could be absorbed due to the pair production process.
This effect leads to a change in the spectral shape of observed radiation and it can be used to constrain the EBL density (see e.g. \cite{eblpaper}). 

A possible extended emission from AGN objects was discussed in \cite{aharonian_halo}.
Gamma rays with energies $\gtrsim 100$ TeV can be easily produced in hadronic models of AGNs. 
At those energies $\gamma$-rays will be strongly absorbed in a pair production process on CMB photons already relatively close to the source.
For an Extra-galactic Magnetic Field (EGMF) strength of $\sim 10^{-9}\mathrm{G}$, the produced $e^+e^-$ pairs will be almost immediately isotropized and efficiently cooled down in an inverse Compton process. 
This way a physical halo (of the radius $>1$Mpc) of $e^+e^-$ pairs producing VHE $\gamma$-rays by Inverse Compton scattering can be created in the vicinity of the source.
This extended emission is an input for the second step, where cascading on EBL photons takes place in the intergalactic space between the source and the observer. 
Due to a relatively strong magnetic field, produced $e^+e^-$ pairs are scattered away from the observer field of view.



According to most of the theoretical models a considerable fraction of the intergalactic space has to be filled with a very weak EGMF (\cite{neronov2}).

Unlike \cite{aharonian_halo} the authors in \cite{neronov2}, \cite{neronov}, \cite{kachelriess} are considering cascading in the much weaker EGMF of $\lesssim 10^{-12}\mathrm{G}$.
Because of small deflection angles in weak magnetic fields, the $e^+e^-$ pairs created in the intergalactic space are not isotropized. 
In an inverse Compton process electron-positron pairs will produce secondary gammas with slightly different direction with respect to the primary photons.
An electromagnetic cascade can develop if the optical depths for those secondary photons are still large enough .
Secondary photons produced in those cascades can be deflected and redirected into the field of view of the observer.
Also this process can lead to an apparent extended emission around the point-like source.

The $\gamma$-ray emission from blazars is strongly variable.
The variability timescales could be as short as a few minutes. 
In the case of an extended emission, depending on the distance from the source on which the pair production process occurred, the geometrical path-length transversed by $\gamma$-rays could be rather different.
Due to this geometrical effect, the original short time variability will be smeared out to much longer time scales ($\sim10^5$ years \cite{neronov}).

So far no extended emission around AGN object has been found (see e.g. \cite{hegra_halo}).

MAGIC is an Imaging Atmospheric Cerenkov Telescope located on Canary island of La Palma at Roque de los Muchachos Observatory (2200 m.a.s.l).
Due to the large size of the mirror dish and improved light sensors, it has the lowest energy threshold among all IACTs.
Recently the MAGIC project has been upgraded by adding a second telescope at 85 m distance for working in a coincidence (stereo) mode.

\section{Analysis method}
The image of an every given event can be parametrized by using the so-called Hillas parameters \cite{hillas}. 
The angular distance between the center of gravity of the image and the shower direction (so called DISP) is correlated with geometrical and timing properties of the image. 
Thanks to this the arrival direction for every event can be estimated. 
The distribution of the squared distance between the estimated and the true source position, (the $\theta^2$) is narrow and has a peak at $\theta=0$ for a point-like $\gamma$-ray source. 
In case of an extended source this distribution is broader.

For a single telescope the DISP is parameterized (see \cite{disp_paper}).  
In a first approximation DISP is proportional to the ellipticity ($1-Width/Length$) of the image. 
By including the dependence on the image parameter $Size$  and it's possible truncation because of the camera edge effects, one improves the precision of DISP. 

\subsection{The novel Random Forest DISP method}
For this study we developed a novel method for the DISP estimation. 
As shown in \cite{timepaper} for a given image the time gradient along its main axis is strongly correlated with the impact parameter of the parent shower. 
Since for a given zenith angle and a $\gamma$-ray energy the DISP is a simple function of the impact parameter, one could expected that by using fast timing properties, one will improve the DISP estimation, and as a result also the angular resolution. 

\begin{figure}
  \centering
  \includegraphics[width=8cm]{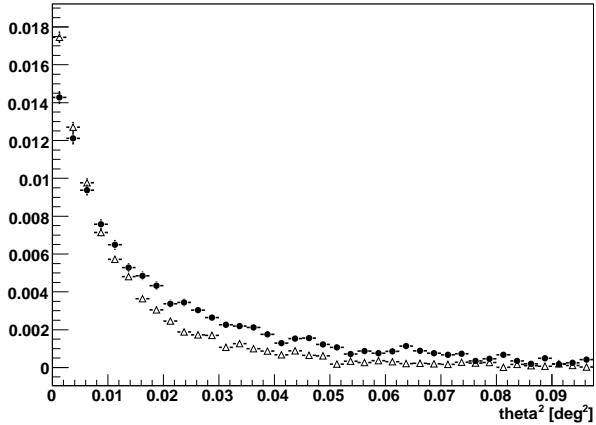}
  \caption{A comparison of $\theta^2$ distributions obtained with parametrized DISP value (circles), and with a RF DISP (triangles) for the Crab Nebula data.
}\label{fig_rf_vs_par}
\end{figure}

To combine information from both geometrical and timing properties of the image in the most efficient way, we used multidimensional decision trees - the Random Forest (RF) method. 
It is widely used for the gamma/hadron separation and the energy estimation \cite{RF}. 
A comparison of the novel RF method (hereafter RF DISP) and of the standard parametrized DISP is presented in fig.~\ref{fig_rf_vs_par}.
The RF DISP provides a substantially narrower $\theta^2$ distribution.
This improves the angular resolution by $\sim 20-30\%$ and thus enhances the telescope performance for the search of an extended emission. 

The shape and the width of a $\theta^2$ distribution for a point-like source depends on many factors. 
The most important is the energy of the shower. 
For higher energy showers, due to large number of particles in the shower maximum the relative internal fluctuations are smaller, and the resulting image has better defined parameters.
Also the large $Size$ (the total measured charge of the image) improves the precision of the reconstruction of the shower direction. 
In our analysis we used only showers with $Size>400 $photoelectrons, that provide a relatively high precision of the $\theta^2$ determination.
This leads to an energy threshold (defined as a peak of the Monte Carlo energy distribution) of 300 GeV.

\begin{figure}
  \centering
  \includegraphics[width=8cm]{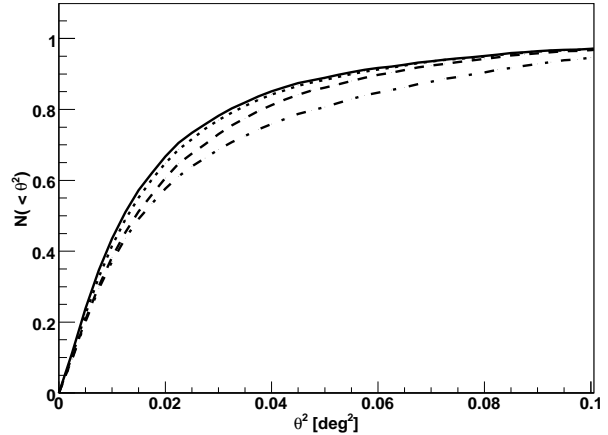}
  \caption{
Monte Carlo cumulative $\theta^2$ distributions for a source with 80\% point-like and 20\% extended emission. 
Characteristic radius of the extension is equal to $0.1^\circ$ (dotted), $0.2^\circ$ (dashed) or $0.3^\circ$ (dot-dashed).
The extended part of the emission is simulated as a flat disc ($dN/d\theta^2=\mathrm{const}$).
Purely point-like source is shown with solid line.
A random misspointing up to 0.03 deg has been included in the simulations.}
  \label{fig_th2cum}
\end{figure}

Let us consider a situation when the excess observed from a hypothetical source is a mixture of a point source and a small addition of extended emission with a given profile.
The cumulative $\theta^2$ distributions for a point-like and extended sources are shown in fig.~\ref{fig_th2cum}.
Using this figure, the telescope PSF (point spread function) defined as 41\% containment radius (equivalent to one standard deviation of a two dimensional gaussian distribution) of the reconstructed direction of $\gamma$-rays for the used analysis energy threshold is estimated to be $\lesssim 0.1^\circ$.

If the characteristic extension $\theta_{ext}$ is smaller than the telescope's PSF, the distributions of a purely point-like and a partially extended source are very similar. 
For larger extensions clear differences between distribution shapes can be seen (see fig.~\ref{fig_th2cum}).

To investigate possible extended emission from blazars we adopted a method similar to the one used in \cite{hegra_halo}. We calculated the ratio of event rates in two $\theta$ ranges: 
$f= \frac{N(\theta_1<\theta<\theta_2)}{N(0<\theta<\theta_1)},$
and compared them with the ratio calculated for a point-like source.

The $\theta_1$ and $\theta_2$ values are calculated with the help of Monte Carlo simulations searching for the most significant difference between purely point source and a source with an extended emission for a given size and profile of extended emission. 

To minimize the systematic errors we used the Crab Nebula as a model of a point-like source. 
This can be done because the mean spectral slopes of Mrk 501 and Mrk421 in the considered data sample are rather similar with that of the Crab Nebula. 

For every considered flux, radius and profile of the extended emission by using MC simulations we calculated the value of $f$ and of the corresponding significance of the extension. 

\subsection{Misspointing}
One of the factors which limit IACTs ability to distinguish an extended from a point-like source is the misspointing of the telescope. 
Using a strong source like Crab or Mrk 421 one can estimate the source position and then calculate the misspointing as a difference between the true and the estimated positions.
\begin{figure}
  \centering
  \includegraphics[width=8cm]{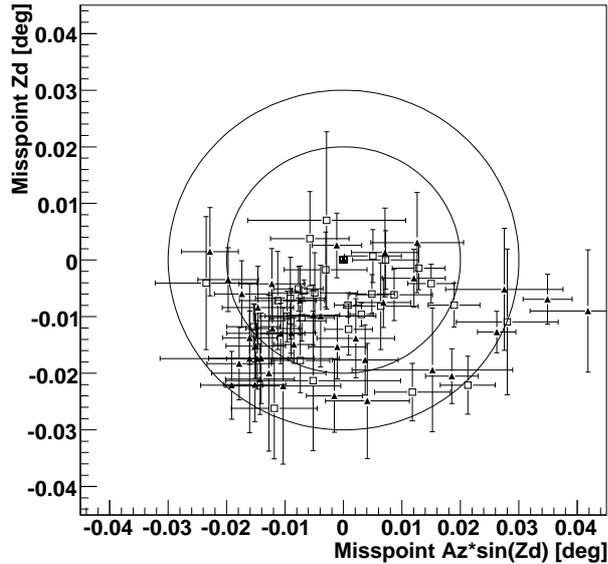}
  \caption{Difference between the true and the estimated source positions for Mrk421 (squares) and Crab Nebula (triangles). 
Circles with radii $0.02^{\circ}$ and $0.03^\circ$ show the characteristic misspointing scale}
  \label{fig_misspoint}
\end{figure}
As it is shown in fig.~\ref{fig_misspoint} the characteristic misspointing scale is  $\sim 0.02-0.03^\circ$
This value is well below the angular resolution of the telescope and the assumed extensions of the halos.

\section{Data sample}

All data presented in this paper has been taken with a single telescope.
To minimize systematic errors, only data taken at low ($<30^\circ$) zenith angle were used. 
Data from two different blazars have been analyzed. 

Data from Mrk~501 have been taken in April/May 2008 in the so called ON/OFF mode (source is in the center of the camera). 
After quality cuts 26h of ON data were obtained. 
Estimation of the background is performed using 50h of OFF data. 
Mrk~501 in the above-mentioned time period was in a rather low state (mean flux in the sample $\sim 15\%$ C.U. (Crab unit))

Mrk~421 data are from the period Dec 2007 - Jan 2009. 
After quality cuts 32h of data were selected. 
Those data were taken in the so called wobble mode. 
Source position was shifted to $0.4^\circ$ from the center of the camera. 
The opposite (with respect to the camera center) position was used for estimating the background. 
Since the background was taken simultaneously with the source data, the systematic errors were small. 
A disadvantage of this approach is the fact that if the source extension is as large as $\sim 0.4^\circ$, the signal and the background regions start overlapping.
This effect, that reduces the sensitivity was taken into account. 
In the analyzed data sample Mrk~421 was in a high state ($\sim 1.4$ C.U.). 

The $\theta^2$ distribution for a point like source was calculated by using the Crab Nebula data.

\section{Results}
\begin{figure}[t]
  \centering
  \includegraphics[width=8cm]{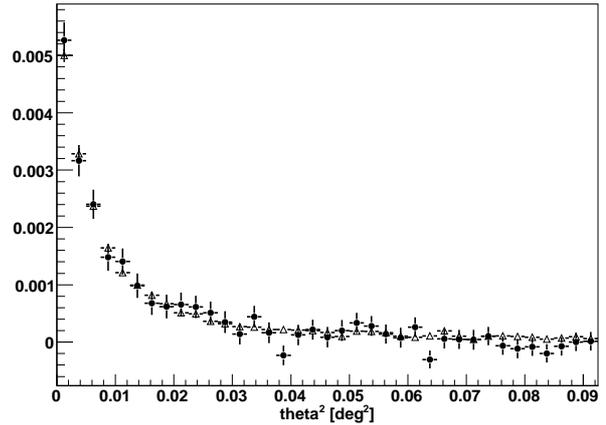}
  \caption{The comparison of the $\theta^2$ distribution for Mrk501 (circles) and a point-like source (triangles, Crab Nebula).}
  \label{fig_mrk501}
\end{figure}

\begin{figure}
  \centering
  \includegraphics[width=8cm]{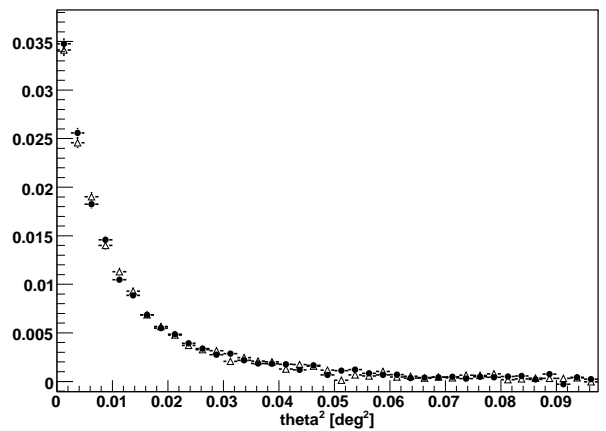}
  \caption{The comparison of the $\theta^2$ distribution for Mrk421 (circles) and a point-like source (triangles, Crab Nebula).}
  \label{fig_mrk421}
\end{figure}

$\theta^2$ distributions obtained for Mrk 501 and Mrk 421 are presented in the fig.~\ref{fig_mrk501} and fig.~\ref{fig_mrk421}.
For both sources the $\theta^2$ distributions match with a corresponding point-like distribution.
Using the first 12 bins, which contain most of the excess events we calculated $\chi^2/n_{dof}$=10.7/11 (for Mrk~421) and 3.8/11 (for Mrk~501). 

\begin{figure*}[t]
  \centering
  \includegraphics[width=8cm]{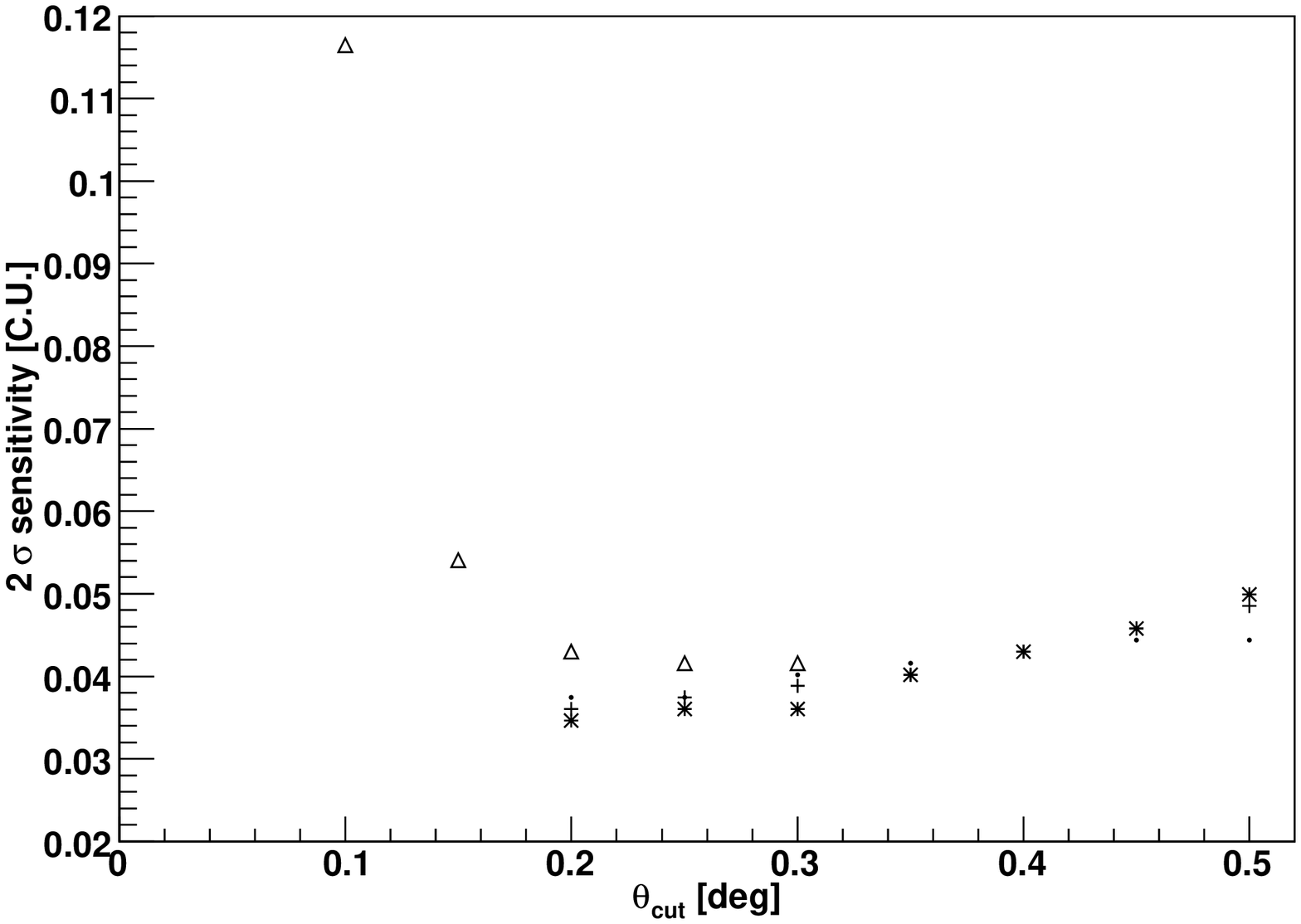}
  \includegraphics[width=8cm]{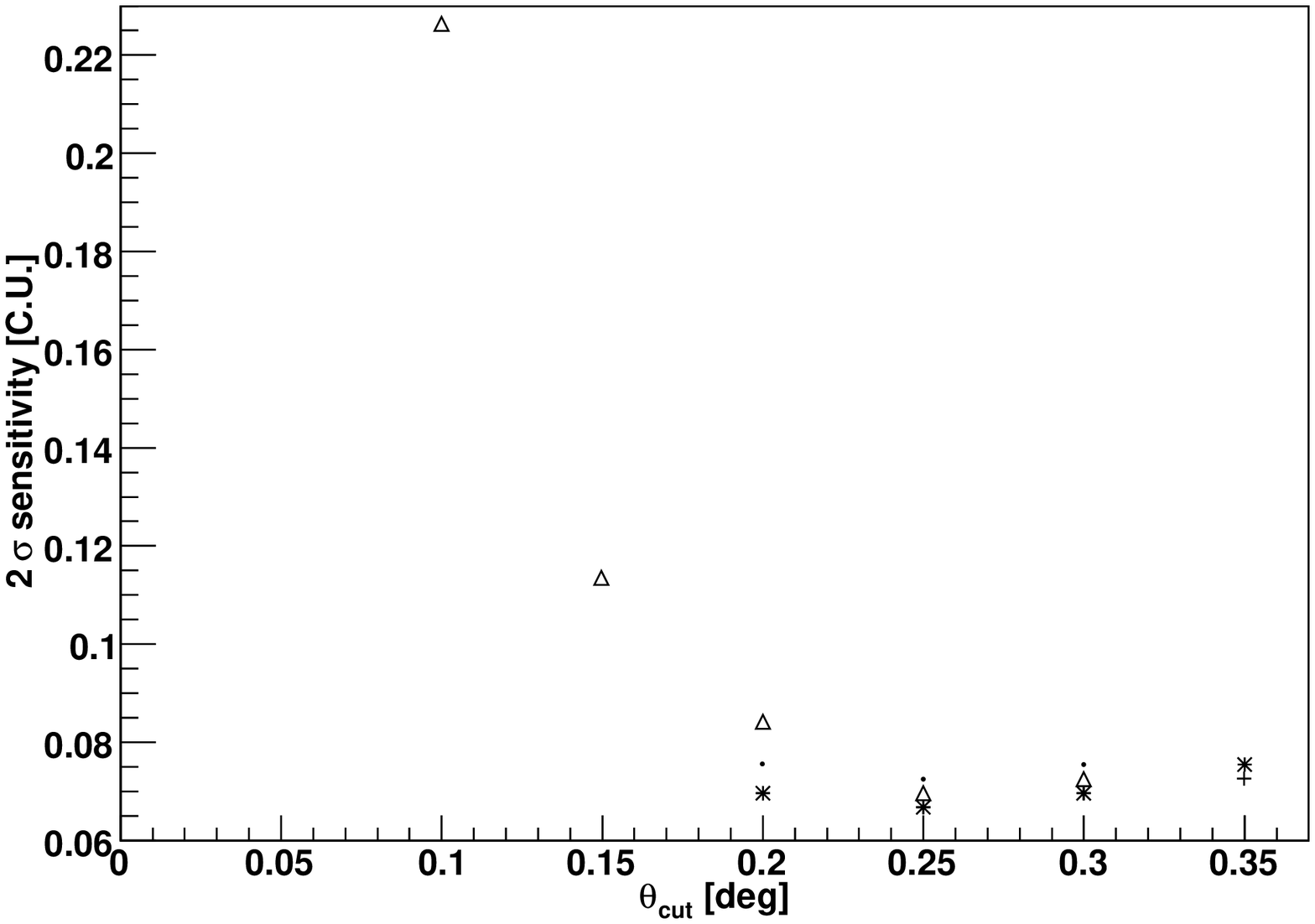}
  \caption{An upper limit on the flux of the extended emission from Mrk 501 (left figure) and Mrk 421 (right figure) in the Crab units for different source profiles and extensions $dN/d\theta \propto \theta^\beta$:
    $0^\circ < \theta < \theta_{cut}$: flat ($\beta=1$) (open circles), $0.1^\circ < \theta < \theta_{cut}$: $\beta=0$ (stars), $\beta=-1$ (crosses), $\beta=-2$ (dots).}
  \label{fig_uplim}
\end{figure*}

An upper limit on the flux of the extended emission calculated for different extension radii and profiles is shown in fig~\ref{fig_uplim}. 

\section{Conclusions}
We have developed a novel method based of the Random Forest for the estimation of the source position.
This method improves the angular resolution by $\sim 20-30\%$. 

For the used data samples stemming from 26h of observations of Mrk~501 and 32h of Mrk~421 no extended emission has been detected around these sources.
The study showed that if there is an extended emission around Mrk 501, then its flux is $<$ 4\% C.U. (see fig.~\ref{fig_uplim} left), for the analysis energy threshold 300 GeV. 
In the case of Mrk 421, since the source was in a high emission state, the search for the halo was less sensitive, because point-like emission creates additional background. 
The constrain on the extended emission flux from Mrk~421 is $< 7\%$ C.U. (see fig.~\ref{fig_uplim} right).

With the second telescope, the MAGIC system can operate in the stereo mode. 
By combining simultaneous information from both telescopes, the angular resolution will be significantly improved. 
This will allow us to perform a search for the AGN halo with a better sensitivity.

A future $\gamma$-ray telescope project - the Cerenkov Telescope Array is aiming for a much better angular resolution ($\sim$2 arc min).
In this case lower halo extensions can be investigated with an improved sensitivity.

\section*{Acknowledgements}
\noindent {\small We would like to thank A. Neronov and D. Semikoz for drawing our attention to this interesting subject.}


\begin{thebibliography}{99}
  \bibitem{eblpaper} Mazin, D., Raue, M. 2007, A\&A, 471, 439M	
  \bibitem{aharonian_halo} Aharonian, F. A., Coppi, P.S., V\"olk H.J., 1993, ApJ, 423, L5
  \bibitem{neronov2} Elyiv, A., Neronov, A., Semikoz, D., arXiv:0903.3649
  \bibitem{neronov} Neronov A., Semikoz D.V., 2007, JETP Lett. 85, 579 
  \bibitem{kachelriess} Dolag, K., Kachelriess, M., Ostapchenko, S., Tomas, R., arXiv:0903.2842
  \bibitem{hegra_halo} Aharonian, F. A., et al. 2001, A\&A, 366, 746
  \bibitem{hillas} Hillas, A. M., 1985, Proc. 19th ICRC, La Jolla 3, 445
  \bibitem{disp_paper} Domingo-Santamaria, E., Flix, J., Scalzotto, V., Wittek, W., Rico, J., 2005, Proc. 29th ICRC, Pune, 101
  \bibitem{timepaper} E. Aliu et al., Astropart. Phys. 2009, 30,  293
  \bibitem{RF} Albert, J. et al., 2008, NIM A, 588, 424

  \end{thebibliography}
\end{document}